\documentclass{article}
\usepackage[english]{babel}
\usepackage[section]{placeins} 
\usepackage[letterpaper,top=2cm,bottom=2cm,left=3cm,right=3cm,marginparwidth=1.75cm]{geometry}
\usepackage[colorlinks=true, allcolors=blue]{hyperref}
\usepackage{apacite}
\usepackage{amsmath}
\usepackage{graphicx}
\usepackage{comment}
\usepackage{authblk}

\title{Deconstructing written rules and hierarchy in peer produced software communities}

\author{Mahasweta Chakraborti\thanks{Corresponding Author}, Qiankun Zhong, Beril Bulat, Seth Frey}
\affil{University of California, Davis, USA}
\author{Anamika Sen}
\affil{UMass Amherst, Amherst, MA, USA}
\begin{document}
\maketitle

\section{Introduction}

\begin{flushleft}
The internet has changed how we build organizations, and poses fundamental challenges to basic organizational scholarship \cite{kollock1999economies}. Peer production in particular has changed how we view hierarchy, empowerment, motivation, and authority. And yet, other scholarship has challenged the revolutionary view of peer production, demonstrating that despite a rhetoric of flatness and democratic access, it is vulnerable to the emergence of the power structures typical of traditional organizations \cite{10.1111/jcom.12082}.
\end{flushleft}
\begin{flushleft}
This tension is particularly clear in the case of Open Source Software development (OSS). OSS is a multi billion dollar industry and the backbone of major enterprise level software products and services \cite{schweik2012internet}. OSS projects are largely volunteer driven initiatives where fleeting developers join, cooperate or cease contributing of their own volition. Thus, there are less bindings and contracts when compared against the monolithic software companies comprising of dedicated paid employees and established hierarchies \cite{HERTEL20031159}.
On the other hand, these OSS projects are often mentored and supported by OSS organizations in surprisingly hierarchical structures, in terms of outreach, licensing, infrastructure and other logistics \cite{10.1007/978-3-319-57735-7_2}.
\end{flushleft}
\begin{flushleft}
%Several OSS organizations have instituted their own incubator programs for new projects seeking admission. 
Established in 1999, the Apache Software Foundation is one of the most recognizable and well documented OSS organizations, which currently houses over 350 projects. The ASF is a strong proponent of the core OSS philosophy of free, open participation, resists authoritative hierarchy \cite{apache} and encourages community building and consensus-based self governance in projects. 

\end{flushleft}

\begin{flushleft}
We employ recent advances in computational institutional analysis and NLP to investigate the systems of authority that are reflected in the written policy documents of the ASF. Our study to decipher the effective similarities or departures of the ASF model from conventional software companies reveals evidence of both flat and bureaucratic governance in a peer production set up, suggesting a complicated relationship between business-based theories of administrative hierarchy and foundational principles of the OSS movement.
    
\end{flushleft}

\section{Literature Review}
\begin{flushleft}
OSS has been described as peer-produced public good generation through generalized exchange where developers are set to mutually benefit as contributors as well as users \cite{kollock1999economies}.In the past twenty years, the emergence of peer production has drawn much attention in communication \cite{meng2013commons,10.1111/jcom.12082}, economics\cite{benkler2006commons, birkinbine2020political}, and political science\cite{johnson2004accountable}. While the early scholars are fascinated about the idea of cooperation and decentralization in the process of producing and distributing digital information \cite{benkler2006commons}, later scholars started to consider how much real difference peer production organizations have with the conventional model of authority and bureaucracy \cite{kreiss2011limits} based on the increasing number of empirical work  \cite<e.g.,>{bennett2014organization}.
\end{flushleft}
\begin{flushleft}
The early optimists focus mostly on the fundamental changes this new form of information production brings to the power structure and cooperation norms in the organization. Specifically, Benkler\citeyear{benkler2006commons} argues that reduced cost in information production and distribution creates market incentives for individuals to get involved, which motivates a strong norm of cooperation. At the same time, the process of peer production provides the contributor with a sense of belonging and bonding \cite{benkler2006commons}. Thus, peer production is expected to have a structural effect on organizations and societies towards a more egalitarian and decentralized model compared to the conventional Weberian models \cite<e.g.,>{bauwens2008ladder,de2006publicly}.
\end{flushleft}
\begin{flushleft}
For some others, peer production projects closely approximate conventional organizational forms. This perspective has a long tradition \cite{weber1964theory}, and views top-down structures as a requirement for efficiency and consistency. Some argue that bureaucratic structures are essentially flexible instruments for achieving organizational goals, a means for power centralization \cite{perrow1986complex}. But power is an inherently unstable concept, which must be understood within context-specific terms \cite{perrow1986complex}. How power is structured within an organization, therefore, depends on how task specification relates to the hierarchy for achieving organizational goals. There is evidence that supports this flexible view of organizational structures in the online context. Shaw and Hill \citeyear{10.1111/jcom.12082} found, for instance, that large-scale peer production projects, such as wikis, might exhibit an oligarchic structure rather than democratic, where authority positions are monopolized among a few early users. And, as the number of voluntary contributors grows, power gets increasingly centralized, despite missing an established bureaucratic structure. 
\end{flushleft}
\begin{flushleft}
An important and often overlooked source of insight into authority in organizations are it's written policies, which collectively specify the governance institution that it operates as. Institutions comprise of basic units in the form of rules, norms and strategies (Institutional Statements) which seek to define and regulate behaviour of individuals who participate in it\cite{CrawfordOstrom1995}. To analyze ASF policy documents for finer-grained insight into the administrative structure of OSS institutions, we use a tool called the institutional grammar (ABDICO)  \cite{frantz2021institutional}, which was formulated as means towards systematic analysis of institutional configuration, structural emergence and policy outcomes from policy texts. ABDICO sets up a protocol to decompose statements of institutional structure, such as rules and norms, into syntactic components.  The components identified under the framework include (i) Agent(A) or the actor who the institutional statement concerns (called the Attribute in the ABDICO literature) (ii)  Object(B) animate or inanimate, receives the action from the actor. (iii) Deontics(D) or modals indicate the extent to which the statement needs to be observed/enforced (ex. must/may/must not) (iv) Aim(A) or the action performed by Attribute (v) Context(C), introduced as a supplement to 'Conditions' in the original IAD. Encompasses the conditions and constraints under which action needs to be performed (vi) Or else (O) states the penalty for non observance of the Institutional Statement. With these components, we extend familiar organizational understandings to hierarchy and authority to include quantitative representations of those constructs as they are encoded in organizations' written institutional structures.
\end{flushleft}
\begin{flushleft}
It is still not clear if open source software development projects are susceptible to conventional power structures, or if they actually present a democratic case of peer production as a voluntary organization.  This work represents an attempt to explore this question:
\end{flushleft}
\begin{quote}
    \emph{RQ: What are the interactional dynamics of project mentorship in OSS projects, and how are they different from conventional organizations, if at all?}
\end{quote}
% \section{Related Work} 

% \section{Related Work}
%\section{Institutional Grammar Analysis}

\section{Methodology}

\subsection{Data}
Our data set comprises of 328 policy statements that were hand coded by an expert annotator through a comprehensive analysis of the Apache Incubator Policy, Community Guide, PPMC Guide, Apache Cookbook, Mentor Guide, Graduation Guide, Retirement Guide and the Release Management Guide.

\begin{table}[hbt!]

\centering
\begin{tabular}{|p{0.5\linewidth} | p{0.2\linewidth}|}
\hline
\textbf{Policy Document} & \textbf{Policy count} \\ \hline
% Podlings  & ('podlings', 6), ('podling', 3), ('the podling', 1), ('podling releases', 1), ('apache podling name', 1), ('podling web sites', 1), ('the podling votes', 1), ('podlings wishing', 1), ('podling community members', 1), ('podlings who are unsure', 1), ('podling website dir', 1) \\ \hline
 Apache Incubator Policy & 45 \\ \hline
Community Guide & 28 \\ \hline
PPMC Guide & 61 \\ \hline
Apache Cookbook & 58 \\ \hline
Mentor Guide & 36\\ \hline
Graduation Guide & 51 \\ \hline
Retirement Guide & 22 \\ \hline
Release Management Guide & 26 \\ \hline
\end{tabular}
\caption{\label{tab:widgets}ASF policy documents used for analysis in our study and the number of rules in each.}
\end{table}

\subsection{IGextract and Topic Modeling}
\begin{flushleft}
We begin our analysis of ASF  policies by automated extraction of the ABDICO 2.0 components from the statements in the dataset through the IGextract library \cite{rice2021machine}. The IGextract was initially developed for ABDICO codification of food policies \cite{siddiki2014assessing}.  For our current purposes, we extended the IGextract pipeline beyond training and validation, to generalize ABDICO label prediction for new data. This involved generating word level dependency-based features  as well as contextual vectors \cite{manning2014stanford} using BERT\cite{devlin2018bert} across all statements in the annotated training data as well as ASF policies, followed by training and prediction through a dense, linear neural network. Labels were generated for every word in a statement. Contiguous words with the same ABDICO label were grouped together.
\end{flushleft}
\begin{flushleft}
Following ABDICO coding, we apply unsupervised topic modelling on these extracted components in order to group rules by their Attributes, Objects and Deontics. We particularly focus on the these components in order to determine the scope of the rules and norms in the ASF. The clustering employs BertTopic,\cite{grootendorst2020bertopic} which identifies topics and clusters across a set of documents. For each component extracted across all the statements, BertTopic generates contextual vector representations through BERT, which undergo dimensionality reduction through UMAP followed by clustering with HDBSCAN.   BertTopic parameters were selected for localised clustering. 
\end{flushleft}
\subsection{Clustering Agents and Objects}
\begin{flushleft}
The semantic clustering was followed by manual aggregation of associated clusters based on domain understanding of the Apache ecosystem. We then identified each of these super clusters with high level Agents(Authority and Participants) and Objects(Participants, Product, Project Management and Product Management). For details on these clusters see Appendix. 

\begin{figure}[hbt!]
\centering
\includegraphics[width=0.9\textwidth]{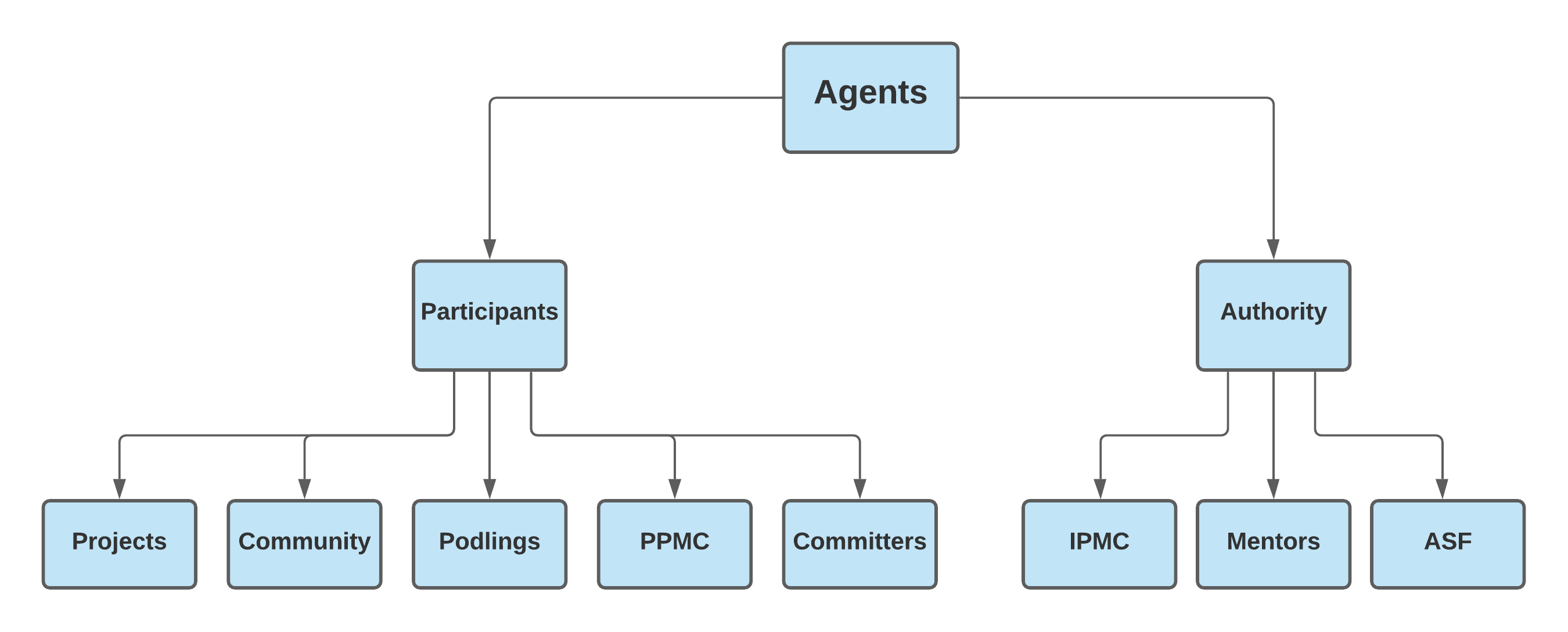}
\caption{\label{fig:attr_levels} Agent clusters in the ASF policies could be identified as either authority figures of participants }
\end{figure}

\begin{figure}[hbt!]
\centering
\includegraphics[width=0.9\textwidth]{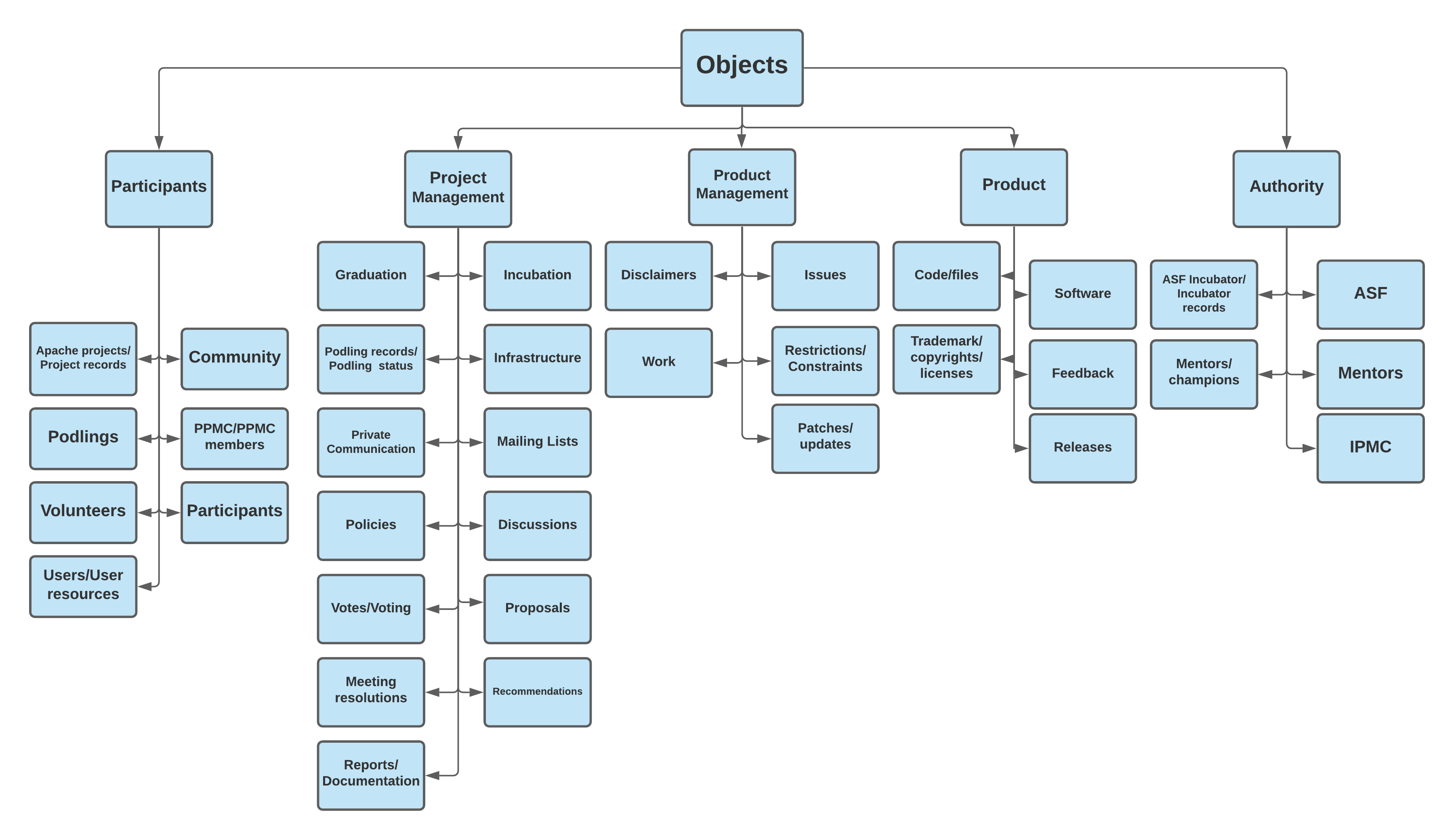}
\caption{\label{fig:obj_levels} Object clusters found in the ASF policies were found to be associated with at least one of Product management, Project management, authority, product or participants }
\end{figure}

\end{flushleft}

% \subsection{How to add Tables}

% Use the table and tabular environments for basic tables --- see Table~\ref{tab:widgets}, for example. For more information, please see this help article on \href{https://www.overleaf.com/learn/latex/tables}{tables}. 

\section{Analysis and Results}

\begin{flushleft}
Despite their democratic potential to decentralize knowledge production and distribution, OSS projects show a similar administrative structure and hierarchical labor division to conventional organizations, with more rules constraining lower-level than higher-level agents (Figure 3 and 4). However, when comparing the content of those institutional constraints, we did support the claim that OSS projects reflect a more egalitarian governance style. An examination of how increasingly harsh rule conditions (deontics) are imposed upon authorities versus participants showed no statistically significant differences ($\chi^2_2 = 0.021$, $p = 0.99$; $\chi^2_2 = 0.38$, $p = 0.83$; $N = 63$), indicating that agents at lower levels, even if they are regulated with more rules, are not regulated with more or less imposing rules: they are regulated in the same way. 
\end{flushleft}

\begin{figure}[hbt!]
\centering
\includegraphics[width=0.9\textwidth]{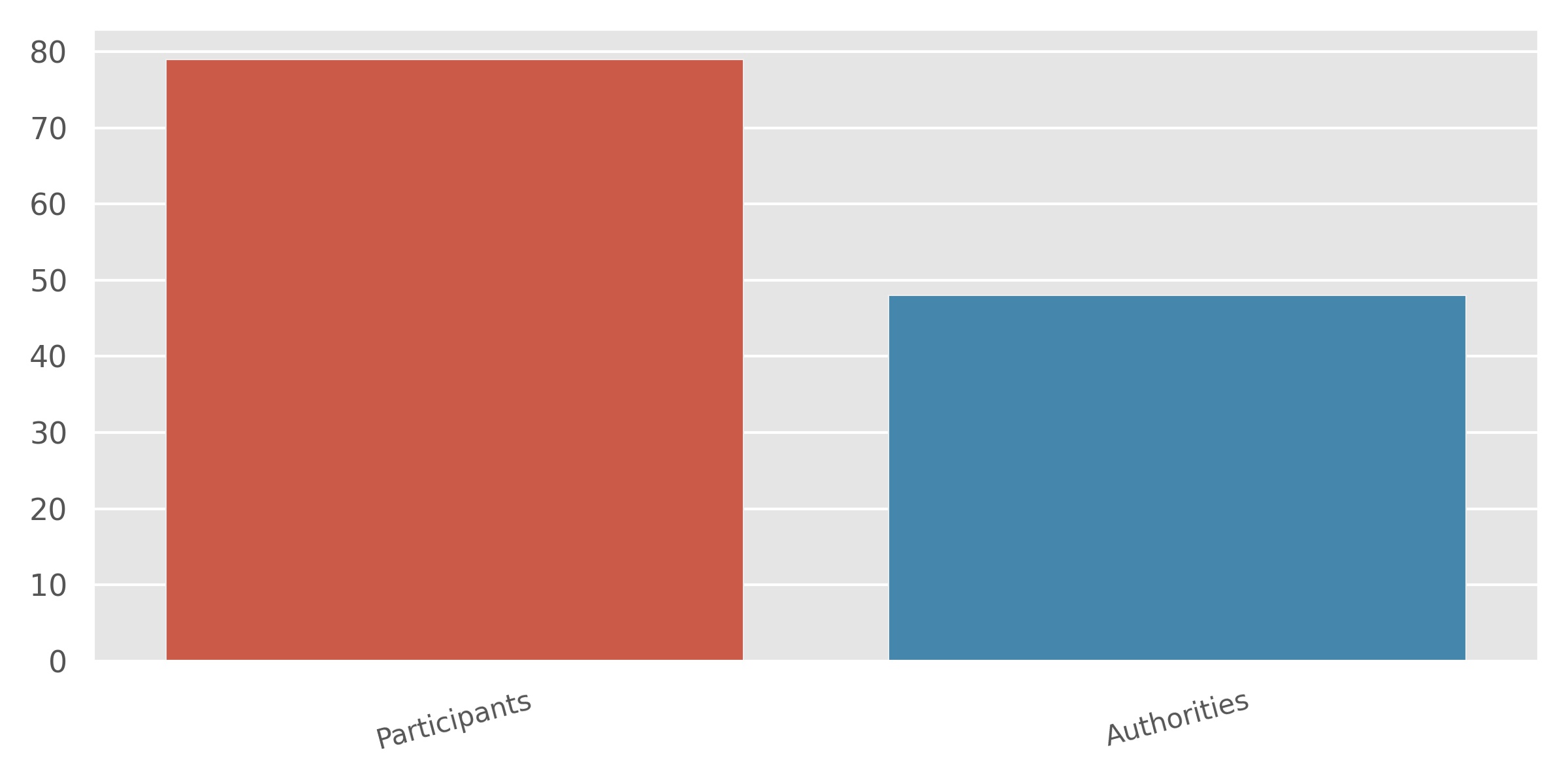}
\caption{\label{fig:attributes}  \textbf{More rules govern volunteer participant behavior than volunteer administrator behavior.} Volunteer contributors in the Apache Software Foundation (ASF) are engaged in peer production around open source software, constrained in part by written institutional statements.  After aggregating the lower-level agents of those sentences by their relative authority into Participants and Authorities, this histogram demonstrates that more institutional statements govern behavior of participants than authorities, consistent with a surprisingly conventional approach to administrative hierarchy. See \textbf{Figure 1} for the mapping of lower-level agents to these two categories. See \textbf{Figure 8} for a histogram of the component clusters of low-level agents.   }
\end{figure}

% A classifier based on advances in natural language processing extracted the components of the institutional statements that constitute ASF’s written institutional structure. 

\begin{flushleft}
We then look into what is being governed and managed in OSS projects by analyzing the grammatical object in institutional statements, including environment (i.e. community) and resources (i.e. code files). We found that most of the agent behaviors were governed with respect to the project management infrastructure or other participants (Figure 4).
\end{flushleft}

\begin{figure}[hbt!]
\centering
\includegraphics[width=0.9\textwidth]{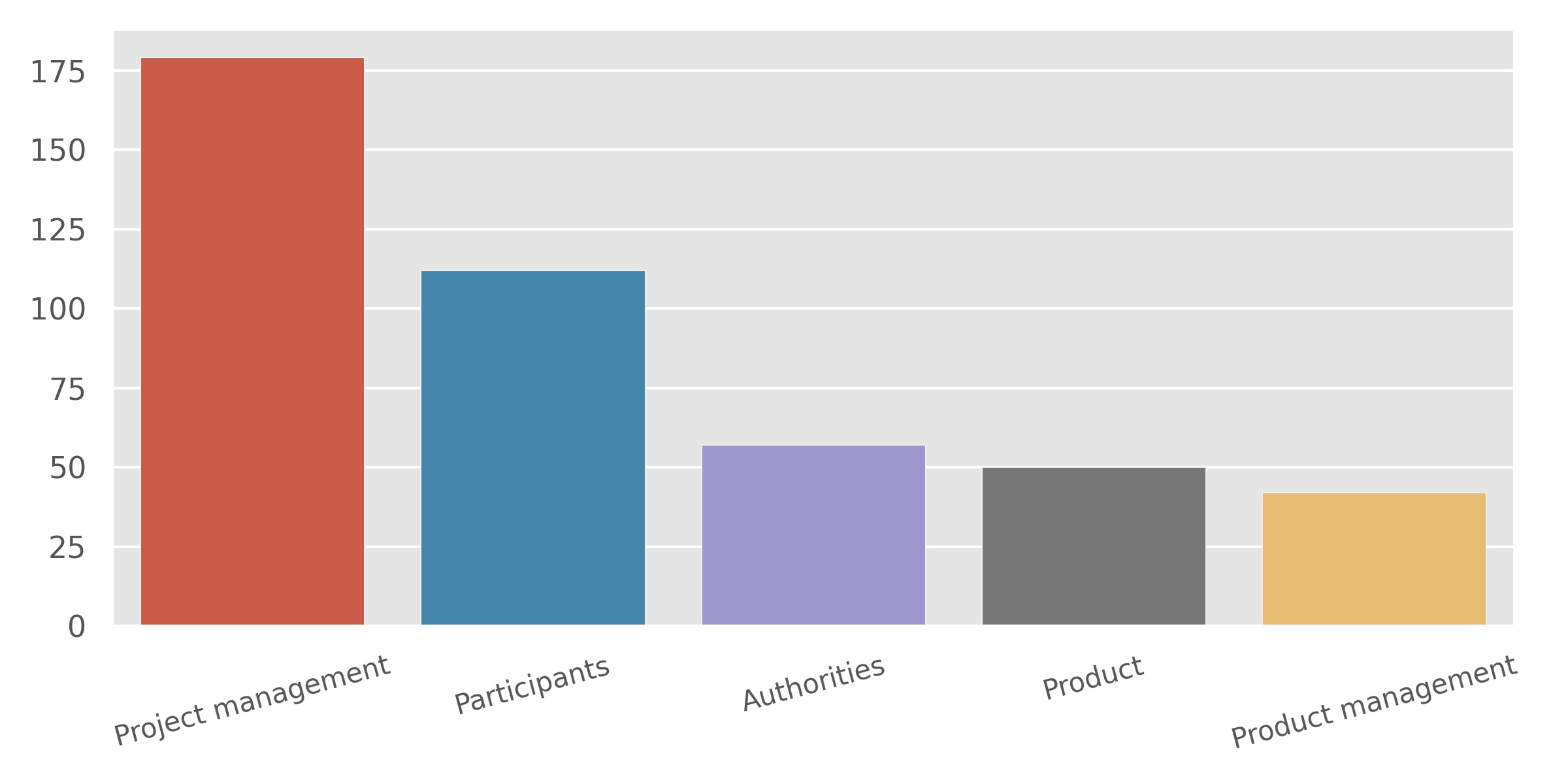}
\caption{\label{fig:objects} \textbf{Agent behavior in the ASF is mostly coordinated around project management and participants.} In the context of the institutional grammar, an institutional statement can define or constrain an agent’s behavior with respect to a grammatical object. The object can be a part of the institution or it’s environment, membership, or resources. According to this histogram, agents behavior is most often governed with respect to a project’s management infrastructure (repositories, mailing lists, voting protocols), or other participants.  Authorities and product management infrastructure are less likely to be the vertex of an institutional statement’s governance effort. See \textbf{Figure 1} for the mapping of lower-level objects to these five categories. See \textbf{Figure 8} for a histogram over unaggregated low-level objects }
\end{figure}
\begin{flushleft}
We also use the deontic component of institutional statements to measure the strength of the requirements prescribed by the rules. For instance, deontics such as “must” or “will” indicate stronger institutional enforcement than “may” or “can”. In our dataset, we found a larger part of the institutional statements governing ASF activity was in this former category (Figure 5). But there was no proportional difference in how they were distributed between participants and authorities (Figure 6a), suggesting that the hierarchies that peer-production organizations implement do, in their details, continue to reflect their egalitarian ideals in balancing the kinds of constraints they impose on their different kinds of volunteer.
\end{flushleft}
\begin{figure}[hbt!]
\centering
\includegraphics[width=0.9\textwidth]{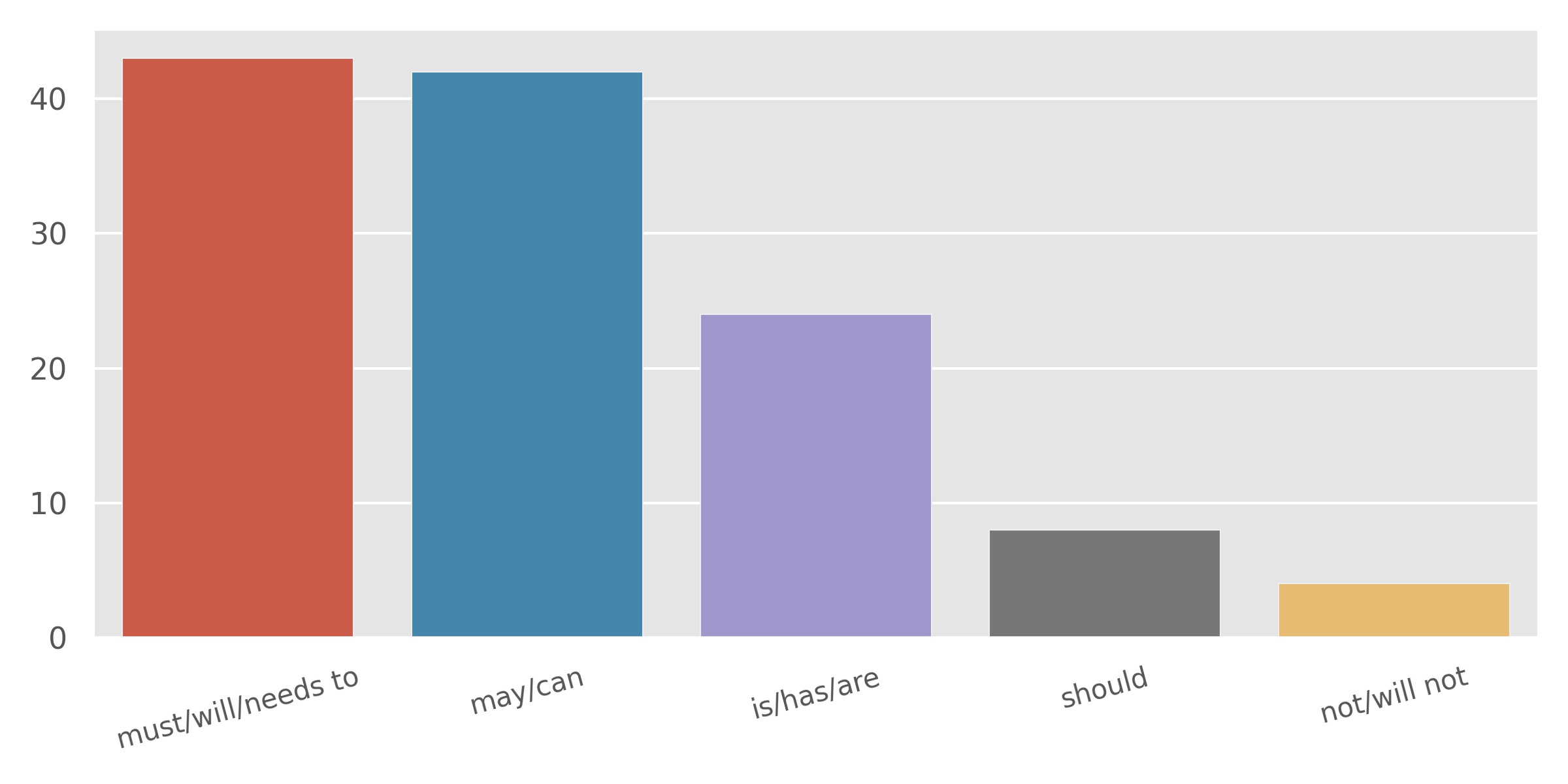}
\caption{\label{fig:deontics} \textbf{Most institutional statements with a deontic component indicated stronger requirements.}  The deontic in an institutional statement can indicate the strength of an institutional prescription, with “must” or “will” defining a stronger rule than “may,” “can,” or even “should.” The histogram indicates that institutional statements governing ASF activities are more likely to rely on strong deontics, and less likely to use proscriptions (will not), verbs for defining institutional state (is/are), or weak deontics (should or may).  This is again consistent with a surprisingly conventional view of administrative hierarchy as it applies to the volunteer labor of OSS developers. }
\end{figure}

\begin{figure}[hbt!]
\centering
\includegraphics[width=0.9\textwidth]{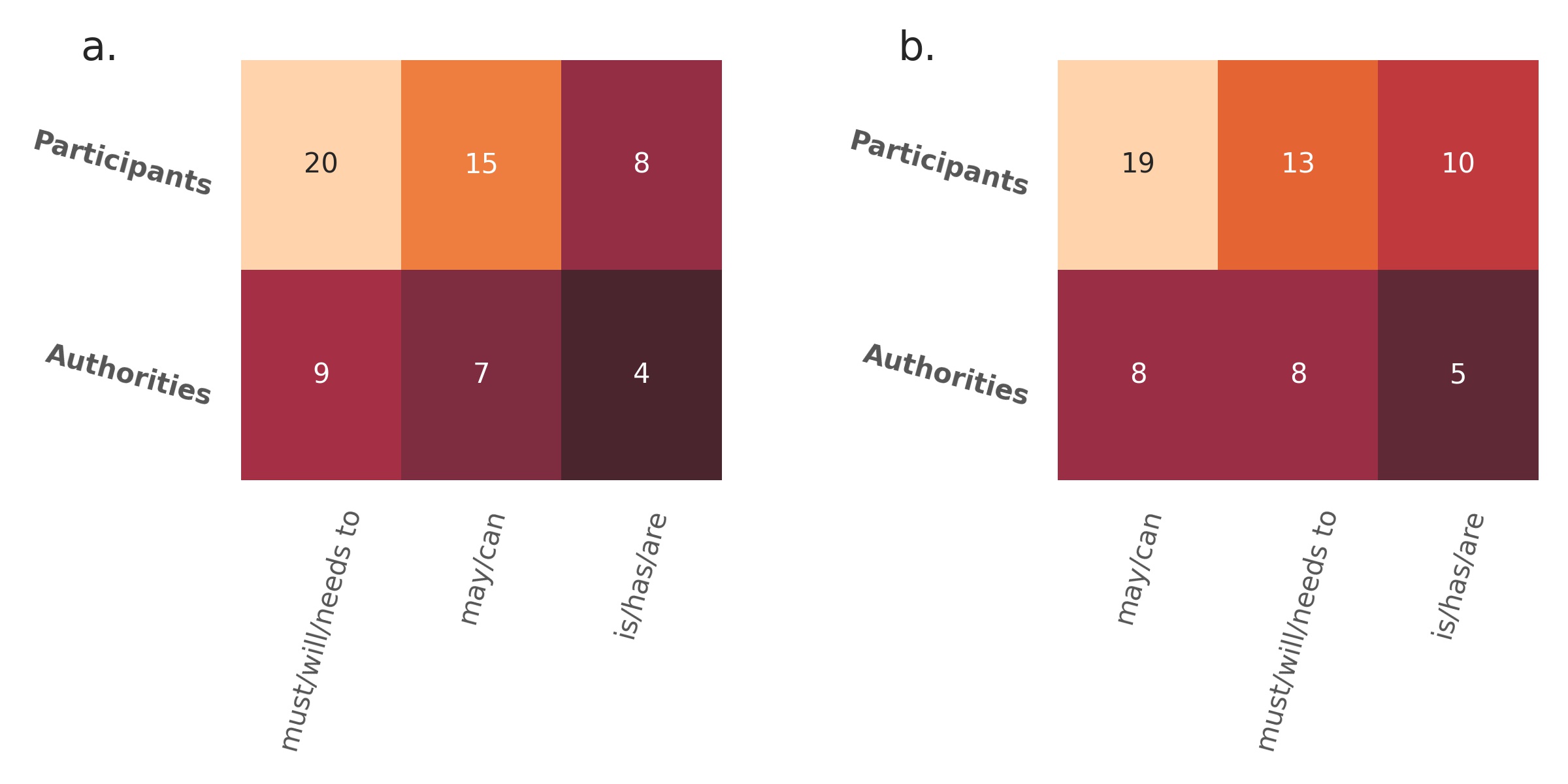}
\caption{\label{fig:crosstabs} \textbf{Participants are governed by the same ratio of strong deontics as administrators, in both the agent and object roles. } These two tables show a “zoomed in” subset of the crosstabulation of deontics against agents and objects, focused on how participants and authorities interact with the top three most common deontics (full tables in \textbf{Figure 6}). Participants and authorities receive the same proportion of strong “must” and “will” than weak “can” or “may” deontics. This suggests that participants and authorities are governed by equally strong constraints. These results are consistent with a conventional understanding of volunteer-driven peer production as expressing a non-traditional understanding of hierarchy and authority. }
\end{figure}

% \begin{table}[hbt!]
% \centering

% \begin{tabular}{|p{0.5\linewidth} | p{0.2\linewidth}|}
% \hline
% Agent & \textbf{\it{p}} \\\hline
% \textbf{Participants} & 0.989 \\ \hline
% \end{tabular}
% \caption{\label{tab:obj} Tabulation of chi-square test for significance of observed differences on the cross tabs between Authorities and all other agents}
% \end{table}

% \begin{comment}

% \begin{table}[hbt!]
% \centering
% \begin{tabular}{|p{0.3\linewidth} |  p{0.3\linewidth}|}
% \hline
% \textbf{Agent} & \\\hline
% \textbf{Participants} & X2 (2, N = 104) = 1.7, p = .05. \\ \hline
% \hline
% \textbf{Object} &  \\\hline
% Product Management & X2 (2, N = 104) = 1.7, p = .05.\\ \hline
% Project Management  & X2 (2, N = 104) = 1.7, p = .05.\\ \hline
% \textbf{Participants} & X2 (2, N = 104) = 1.7, p = .05.\\ \hline
% Product &  X2 (2, N = 104) = 1.7, p = .05.  \\ \hline
% \end{tabular}
% \caption{\label{tab:obj} Tabulation of chi-square test for significance of observed differences on the cross tabs between Authorities and all other objects}
% \end{table}
% \end{comment}

\section{Discussion}
\begin{flushleft}
This research uses novel text data and recent natural language processing tools to explore the administrative hierarchy of a peer production organization. Our results show that although formal rules regulated  participants more than administrators in ASF organizations, the content of those rules was not more restrictive for participants than administrators, indicating that even if participants are subject to a greater number of regulatory constraints, the nature of those constraints is balanced and egalitarian, compared to administrators. This result is consistent with the argument that peer production shifts the power structures and promote more egalitarian administrative systems in making decisions, delegating responsibility and validating individuals’ responsibility. Even with the Weberian lens, we still need to recognize the fundamental difference peer governance mechanisms have compared to conventional centralized organizations.
\end{flushleft}
\begin{flushleft}
Whether this type of administrative structure is efficient for information production and distribution requires further investigation. Built upon current research project, we will analyze the temporal changes in the hierarchy and efficiency, and eventually compare the efficiency of ASF with other more hierarchical peer production organizations in future work.
\end{flushleft}
% \begin{flushleft}

% \end{flushleft}

% We hope you find Overleaf useful, and do take a look at our \href{https://www.overleaf.com/learn}{help library} for more tutorials and user guides! Please also let us know if you have any feedback using the Contact Us link at the bottom of the Overleaf menu --- or use the contact form at \url{https://www.overleaf.com/contact}.

\bibliographystyle{apacite}
\bibliography{main}

\begin{thebibliography}{}

\bibitem [\protect \citeauthoryear {%
ASF%
}{%
ASF%
}{%
{\protect \APACyear {2021}}%
}]{%
apache}
\APACinsertmetastar {%
apache}%
\begin{APACrefauthors}%
ASF.%
\end{APACrefauthors}%
\unskip\
\newblock
\APACrefYearMonthDay{2021}{}{}.
\newblock
\APACrefbtitle {Apache Software Foundation : How it works.} {Apache software
  foundation : How it works.}
\newblock
\APAChowpublished {\url{http://www.apache.org/foundation/how-it-works.html}}.
\PrintBackRefs{\CurrentBib}

\bibitem [\protect \citeauthoryear {%
Bauwens%
}{%
Bauwens%
}{%
{\protect \APACyear {2008}}%
}]{%
bauwens2008ladder}
\APACinsertmetastar {%
bauwens2008ladder}%
\begin{APACrefauthors}%
Bauwens, M.%
\end{APACrefauthors}%
\unskip\
\newblock
\APACrefYearMonthDay{2008}{}{}.
\newblock
{\BBOQ}\APACrefatitle {Ladder of participation: Business models for peer
  production} {Ladder of participation: Business models for peer
  production}.{\BBCQ}
\newblock
\APACjournalVolNumPages{Open Source Business Resource}{}{January 2008}{}.
\PrintBackRefs{\CurrentBib}

\bibitem [\protect \citeauthoryear {%
Benkler%
\ \BBA {} Nissenbaum%
}{%
Benkler%
\ \BBA {} Nissenbaum%
}{%
{\protect \APACyear {2006}}%
}]{%
benkler2006commons}
\APACinsertmetastar {%
benkler2006commons}%
\begin{APACrefauthors}%
Benkler, Y.%
\BCBT {}\ \BBA {} Nissenbaum, H.%
\end{APACrefauthors}%
\unskip\
\newblock
\APACrefYearMonthDay{2006}{}{}.
\newblock
{\BBOQ}\APACrefatitle {Commons-based peer production and virtue} {Commons-based
  peer production and virtue}.{\BBCQ}
\newblock
\APACjournalVolNumPages{Journal of political philosophy}{14}{4}{}.
\PrintBackRefs{\CurrentBib}

\bibitem [\protect \citeauthoryear {%
Bennett%
, Segerberg%
\BCBL {}\ \BBA {} Walker%
}{%
Bennett%
\ \protect \BOthers {.}}{%
{\protect \APACyear {2014}}%
}]{%
bennett2014organization}
\APACinsertmetastar {%
bennett2014organization}%
\begin{APACrefauthors}%
Bennett, W\BPBI L.%
, Segerberg, A.%
\BCBL {}\ \BBA {} Walker, S.%
\end{APACrefauthors}%
\unskip\
\newblock
\APACrefYearMonthDay{2014}{}{}.
\newblock
{\BBOQ}\APACrefatitle {Organization in the crowd: peer production in
  large-scale networked protests} {Organization in the crowd: peer production
  in large-scale networked protests}.{\BBCQ}
\newblock
\APACjournalVolNumPages{Information, Communication \&
  Society}{17}{2}{232--260}.
\PrintBackRefs{\CurrentBib}

\bibitem [\protect \citeauthoryear {%
Birkinbine%
}{%
Birkinbine%
}{%
{\protect \APACyear {2020}}%
}]{%
birkinbine2020political}
\APACinsertmetastar {%
birkinbine2020political}%
\begin{APACrefauthors}%
Birkinbine, B\BPBI J.%
\end{APACrefauthors}%
\unskip\
\newblock
\APACrefYearMonthDay{2020}{}{}.
\newblock
{\BBOQ}\APACrefatitle {Political economy of peer production} {Political economy
  of peer production}.{\BBCQ}
\newblock
\APACjournalVolNumPages{The Handbook of Peer Production}{}{}{33--43}.
\PrintBackRefs{\CurrentBib}

\bibitem [\protect \citeauthoryear {%
Crawford%
\ \BBA {} Ostrom%
}{%
Crawford%
\ \BBA {} Ostrom%
}{%
{\protect \APACyear {1995}}%
}]{%
CrawfordOstrom1995}
\APACinsertmetastar {%
CrawfordOstrom1995}%
\begin{APACrefauthors}%
Crawford, S.%
\BCBT {}\ \BBA {} Ostrom, E.%
\end{APACrefauthors}%
\unskip\
\newblock
\APACrefYearMonthDay{1995}{}{}.
\newblock
{\BBOQ}\APACrefatitle {A Grammar of Institutions} {A grammar of
  institutions}.{\BBCQ}
\newblock
\APACjournalVolNumPages{American Political Science Review}{89}{3}{582--600}.
\PrintBackRefs{\CurrentBib}

\bibitem [\protect \citeauthoryear {%
De~Valk%
\ \BBA {} Martin%
}{%
De~Valk%
\ \BBA {} Martin%
}{%
{\protect \APACyear {2006}}%
}]{%
de2006publicly}
\APACinsertmetastar {%
de2006publicly}%
\begin{APACrefauthors}%
De~Valk, G.%
\BCBT {}\ \BBA {} Martin, B.%
\end{APACrefauthors}%
\unskip\
\newblock
\APACrefYearMonthDay{2006}{}{}.
\newblock
{\BBOQ}\APACrefatitle {Publicly shared intelligence} {Publicly shared
  intelligence}.{\BBCQ}
\newblock

\PrintBackRefs{\CurrentBib}

\bibitem [\protect \citeauthoryear {%
Devlin%
, Chang%
, Lee%
\BCBL {}\ \BBA {} Toutanova%
}{%
Devlin%
\ \protect \BOthers {.}}{%
{\protect \APACyear {2018}}%
}]{%
devlin2018bert}
\APACinsertmetastar {%
devlin2018bert}%
\begin{APACrefauthors}%
Devlin, J.%
, Chang, M\BHBI W.%
, Lee, K.%
\BCBL {}\ \BBA {} Toutanova, K.%
\end{APACrefauthors}%
\unskip\
\newblock
\APACrefYearMonthDay{2018}{}{}.
\newblock
{\BBOQ}\APACrefatitle {Bert: Pre-training of deep bidirectional transformers
  for language understanding} {Bert: Pre-training of deep bidirectional
  transformers for language understanding}.{\BBCQ}
\newblock
\APACjournalVolNumPages{arXiv preprint arXiv:1810.04805}{}{}{}.
\PrintBackRefs{\CurrentBib}

\bibitem [\protect \citeauthoryear {%
Frantz%
\ \BBA {} Siddiki%
}{%
Frantz%
\ \BBA {} Siddiki%
}{%
{\protect \APACyear {2021}}%
}]{%
frantz2021institutional}
\APACinsertmetastar {%
frantz2021institutional}%
\begin{APACrefauthors}%
Frantz, C\BPBI K.%
\BCBT {}\ \BBA {} Siddiki, S.%
\end{APACrefauthors}%
\unskip\
\newblock
\APACrefYearMonthDay{2021}{}{}.
\newblock
{\BBOQ}\APACrefatitle {Institutional Grammar 2.0: A specification for encoding
  and analyzing institutional design} {Institutional grammar 2.0: A
  specification for encoding and analyzing institutional design}.{\BBCQ}
\newblock
\APACjournalVolNumPages{Public Administration}{}{}{}.
\PrintBackRefs{\CurrentBib}

\bibitem [\protect \citeauthoryear {%
Grootendorst%
}{%
Grootendorst%
}{%
{\protect \APACyear {2020}}%
}]{%
grootendorst2020bertopic}
\APACinsertmetastar {%
grootendorst2020bertopic}%
\begin{APACrefauthors}%
Grootendorst, M.%
\end{APACrefauthors}%
\unskip\
\newblock
\APACrefYearMonthDay{2020}{}{}.
\newblock
\APACrefbtitle {BERTopic: Leveraging BERT and c-TF-IDF to create easily
  interpretable topics.} {Bertopic: Leveraging bert and c-tf-idf to create
  easily interpretable topics.}
\newblock
\APACaddressPublisher{}{Zenodo}.
\newblock
\begin{APACrefURL} \url{https://doi.org/10.5281/zenodo.4381785}
  \end{APACrefURL}
\newblock
\begin{APACrefDOI} \doi{10.5281/zenodo.4381785} \end{APACrefDOI}
\PrintBackRefs{\CurrentBib}

\bibitem [\protect \citeauthoryear {%
Hertel%
, Niedner%
\BCBL {}\ \BBA {} Herrmann%
}{%
Hertel%
\ \protect \BOthers {.}}{%
{\protect \APACyear {2003}}%
}]{%
HERTEL20031159}
\APACinsertmetastar {%
HERTEL20031159}%
\begin{APACrefauthors}%
Hertel, G.%
, Niedner, S.%
\BCBL {}\ \BBA {} Herrmann, S.%
\end{APACrefauthors}%
\unskip\
\newblock
\APACrefYearMonthDay{2003}{}{}.
\newblock
{\BBOQ}\APACrefatitle {Motivation of software developers in Open Source
  projects: an Internet-based survey of contributors to the Linux kernel}
  {Motivation of software developers in open source projects: an internet-based
  survey of contributors to the linux kernel}.{\BBCQ}
\newblock
\APACjournalVolNumPages{Research Policy}{32}{7}{1159-1177}.
\newblock
\begin{APACrefURL}
  \url{https://www.sciencedirect.com/science/article/pii/S0048733303000477}
  \end{APACrefURL}
\newblock
\APACrefnote{Open Source Software Development}
\newblock
\begin{APACrefDOI} \doi{https://doi.org/10.1016/S0048-7333(03)00047-7}
  \end{APACrefDOI}
\PrintBackRefs{\CurrentBib}

\bibitem [\protect \citeauthoryear {%
Johnson%
, Crawford%
\BCBL {}\ \BBA {} Palfrey~Jr%
}{%
Johnson%
\ \protect \BOthers {.}}{%
{\protect \APACyear {2004}}%
}]{%
johnson2004accountable}
\APACinsertmetastar {%
johnson2004accountable}%
\begin{APACrefauthors}%
Johnson, D\BPBI R.%
, Crawford, S\BPBI P.%
\BCBL {}\ \BBA {} Palfrey~Jr, J\BPBI G.%
\end{APACrefauthors}%
\unskip\
\newblock
\APACrefYearMonthDay{2004}{}{}.
\newblock
{\BBOQ}\APACrefatitle {The accountable Internet: Peer production of Internet
  governance} {The accountable internet: Peer production of internet
  governance}.{\BBCQ}
\newblock
\APACjournalVolNumPages{Va. JL \& Tech.}{9}{}{1}.
\PrintBackRefs{\CurrentBib}

\bibitem [\protect \citeauthoryear {%
Kollock%
\ \protect \BOthers {.}}{%
Kollock%
\ \protect \BOthers {.}}{%
{\protect \APACyear {1999}}%
}]{%
kollock1999economies}
\APACinsertmetastar {%
kollock1999economies}%
\begin{APACrefauthors}%
Kollock, P.%
\BCBT {}\ \BOthersPeriod {.}
\end{APACrefauthors}%
\unskip\
\newblock
\APACrefYearMonthDay{1999}{}{}.
\newblock
{\BBOQ}\APACrefatitle {The economies of online cooperation: Gifts and public
  goods in cyberspace} {The economies of online cooperation: Gifts and public
  goods in cyberspace}.{\BBCQ}
\newblock
\APACjournalVolNumPages{Communities in cyberspace}{239}{}{}.
\PrintBackRefs{\CurrentBib}

\bibitem [\protect \citeauthoryear {%
Kreiss%
, Finn%
\BCBL {}\ \BBA {} Turner%
}{%
Kreiss%
\ \protect \BOthers {.}}{%
{\protect \APACyear {2011}}%
}]{%
kreiss2011limits}
\APACinsertmetastar {%
kreiss2011limits}%
\begin{APACrefauthors}%
Kreiss, D.%
, Finn, M.%
\BCBL {}\ \BBA {} Turner, F.%
\end{APACrefauthors}%
\unskip\
\newblock
\APACrefYearMonthDay{2011}{}{}.
\newblock
{\BBOQ}\APACrefatitle {The limits of peer production: Some reminders from Max
  Weber for the network society} {The limits of peer production: Some reminders
  from max weber for the network society}.{\BBCQ}
\newblock
\APACjournalVolNumPages{new media \& society}{13}{2}{243--259}.
\PrintBackRefs{\CurrentBib}

\bibitem [\protect \citeauthoryear {%
Lindman%
\ \BBA {} Hammouda%
}{%
Lindman%
\ \BBA {} Hammouda%
}{%
{\protect \APACyear {2017}}%
}]{%
10.1007/978-3-319-57735-7_2}
\APACinsertmetastar {%
10.1007/978-3-319-57735-7_2}%
\begin{APACrefauthors}%
Lindman, J.%
\BCBT {}\ \BBA {} Hammouda, I.%
\end{APACrefauthors}%
\unskip\
\newblock
\APACrefYearMonthDay{2017}{}{}.
\newblock
{\BBOQ}\APACrefatitle {Investigating Relationships Between FLOSS Foundations
  and FLOSS Projects} {Investigating relationships between floss foundations
  and floss projects}.{\BBCQ}
\newblock
\BIn{} F.~Balaguer, R.~Di~Cosmo, A.~Garrido, F.~Kon, G.~Robles\BCBL {}\ \BBA {}
  S.~Zacchiroli\ (\BEDS), \APACrefbtitle {Open Source Systems: Towards Robust
  Practices} {Open source systems: Towards robust practices}\ (\BPGS\ 14--22).
\newblock
\APACaddressPublisher{Cham}{Springer International Publishing}.
\PrintBackRefs{\CurrentBib}

\bibitem [\protect \citeauthoryear {%
Manning%
\ \protect \BOthers {.}}{%
Manning%
\ \protect \BOthers {.}}{%
{\protect \APACyear {2014}}%
}]{%
manning2014stanford}
\APACinsertmetastar {%
manning2014stanford}%
\begin{APACrefauthors}%
Manning, C\BPBI D.%
, Surdeanu, M.%
, Bauer, J.%
, Finkel, J\BPBI R.%
, Bethard, S.%
\BCBL {}\ \BBA {} McClosky, D.%
\end{APACrefauthors}%
\unskip\
\newblock
\APACrefYearMonthDay{2014}{}{}.
\newblock
{\BBOQ}\APACrefatitle {The Stanford CoreNLP natural language processing
  toolkit} {The stanford corenlp natural language processing toolkit}.{\BBCQ}
\newblock
\BIn{} \APACrefbtitle {Proceedings of 52nd annual meeting of the association
  for computational linguistics: system demonstrations} {Proceedings of 52nd
  annual meeting of the association for computational linguistics: system
  demonstrations}\ (\BPGS\ 55--60).
\PrintBackRefs{\CurrentBib}

\bibitem [\protect \citeauthoryear {%
Meng%
\ \BBA {} Wu%
}{%
Meng%
\ \BBA {} Wu%
}{%
{\protect \APACyear {2013}}%
}]{%
meng2013commons}
\APACinsertmetastar {%
meng2013commons}%
\begin{APACrefauthors}%
Meng, B.%
\BCBT {}\ \BBA {} Wu, F.%
\end{APACrefauthors}%
\unskip\
\newblock
\APACrefYearMonthDay{2013}{}{}.
\newblock
{\BBOQ}\APACrefatitle {Commons/commodity: peer production caught in the Web of
  the commercial market} {Commons/commodity: peer production caught in the web
  of the commercial market}.{\BBCQ}
\newblock
\APACjournalVolNumPages{Information, communication \&
  society}{16}{1}{125--145}.
\PrintBackRefs{\CurrentBib}

\bibitem [\protect \citeauthoryear {%
Perrow%
}{%
Perrow%
}{%
{\protect \APACyear {1986}}%
}]{%
perrow1986complex}
\APACinsertmetastar {%
perrow1986complex}%
\begin{APACrefauthors}%
Perrow, C.%
\end{APACrefauthors}%
\unskip\
\newblock
\APACrefYearMonthDay{1986}{}{}.
\newblock
{\BBOQ}\APACrefatitle {Complex organizations} {Complex organizations}.{\BBCQ}
\newblock
\APACjournalVolNumPages{A critical essay (3rd ed.). Glcnview. lL:
  Scott-Forcs}{}{}{}.
\PrintBackRefs{\CurrentBib}

\bibitem [\protect \citeauthoryear {%
Rice%
, Siddiki%
, Frey%
, Kwon%
\BCBL {}\ \BBA {} Sawyer%
}{%
Rice%
\ \protect \BOthers {.}}{%
{\protect \APACyear {2021}}%
}]{%
rice2021machine}
\APACinsertmetastar {%
rice2021machine}%
\begin{APACrefauthors}%
Rice, D.%
, Siddiki, S.%
, Frey, S.%
, Kwon, J\BPBI H.%
\BCBL {}\ \BBA {} Sawyer, A.%
\end{APACrefauthors}%
\unskip\
\newblock
\APACrefYearMonthDay{2021}{}{}.
\newblock
{\BBOQ}\APACrefatitle {Machine coding of policy texts with the Institutional
  Grammar} {Machine coding of policy texts with the institutional
  grammar}.{\BBCQ}
\newblock
\APACjournalVolNumPages{Public Administration}{99}{2}{248--262}.
\PrintBackRefs{\CurrentBib}

\bibitem [\protect \citeauthoryear {%
Schweik%
\ \BBA {} English%
}{%
Schweik%
\ \BBA {} English%
}{%
{\protect \APACyear {2012}}%
}]{%
schweik2012internet}
\APACinsertmetastar {%
schweik2012internet}%
\begin{APACrefauthors}%
Schweik, C\BPBI M.%
\BCBT {}\ \BBA {} English, R\BPBI C.%
\end{APACrefauthors}%
\unskip\
\newblock
\APACrefYear{2012}.
\newblock
\APACrefbtitle {Internet success: a study of open-source software commons}
  {Internet success: a study of open-source software commons}.
\newblock
\APACaddressPublisher{}{MIT Press}.
\PrintBackRefs{\CurrentBib}

\bibitem [\protect \citeauthoryear {%
Shaw%
\ \BBA {} Hill%
}{%
Shaw%
\ \BBA {} Hill%
}{%
{\protect \APACyear {2014}}%
}]{%
10.1111/jcom.12082}
\APACinsertmetastar {%
10.1111/jcom.12082}%
\begin{APACrefauthors}%
Shaw, A.%
\BCBT {}\ \BBA {} Hill, B\BPBI M.%
\end{APACrefauthors}%
\unskip\
\newblock
\APACrefYearMonthDay{2014}{03}{}.
\newblock
{\BBOQ}\APACrefatitle {{Laboratories of Oligarchy? How the Iron Law Extends to
  Peer Production}} {{Laboratories of Oligarchy? How the Iron Law Extends to
  Peer Production}}.{\BBCQ}
\newblock
\APACjournalVolNumPages{Journal of Communication}{64}{2}{215-238}.
\newblock
\begin{APACrefURL} \url{https://doi.org/10.1111/jcom.12082} \end{APACrefURL}
\newblock
\begin{APACrefDOI} \doi{10.1111/jcom.12082} \end{APACrefDOI}
\PrintBackRefs{\CurrentBib}

\bibitem [\protect \citeauthoryear {%
Siddiki%
}{%
Siddiki%
}{%
{\protect \APACyear {2014}}%
}]{%
siddiki2014assessing}
\APACinsertmetastar {%
siddiki2014assessing}%
\begin{APACrefauthors}%
Siddiki, S.%
\end{APACrefauthors}%
\unskip\
\newblock
\APACrefYearMonthDay{2014}{}{}.
\newblock
{\BBOQ}\APACrefatitle {Assessing Policy Design and Interpretation: An
  Institutions-Based Analysis in the Context of Aquaculture in F lorida and V
  irginia, U nited S tates} {Assessing policy design and interpretation: An
  institutions-based analysis in the context of aquaculture in f lorida and v
  irginia, u nited s tates}.{\BBCQ}
\newblock
\APACjournalVolNumPages{Review of Policy Research}{31}{4}{281--303}.
\PrintBackRefs{\CurrentBib}

\bibitem [\protect \citeauthoryear {%
Weber%
}{%
Weber%
}{%
{\protect \APACyear {1964}}%
}]{%
weber1964theory}
\APACinsertmetastar {%
weber1964theory}%
\begin{APACrefauthors}%
Weber, M.%
\end{APACrefauthors}%
\unskip\
\newblock
\APACrefYear{1964}.
\newblock
\APACrefbtitle {The Theory of Social and Economic Organization: Transl. by AM
  Henderson and Talcott Parsons} {The theory of social and economic
  organization: Transl. by am henderson and talcott parsons}.
\newblock
\APACaddressPublisher{}{Free Press}.
\PrintBackRefs{\CurrentBib}

\end{thebibliography}
\appendix

% \begin{table}[hbt!]

% \centering
% \begin{tabular}{|c|c|}
% \hline
% Item & Quantity \\\hline
% Widgets & 42 \\ \hline
% Gadgets & 13 \\ \hline
% \end{tabular}
% \caption{\label{tab:widgets}An example table.}
% \end{table}

\section{Appendix}

\begin{figure}[hbt!]
\centering
\includegraphics[width=0.9\textwidth]{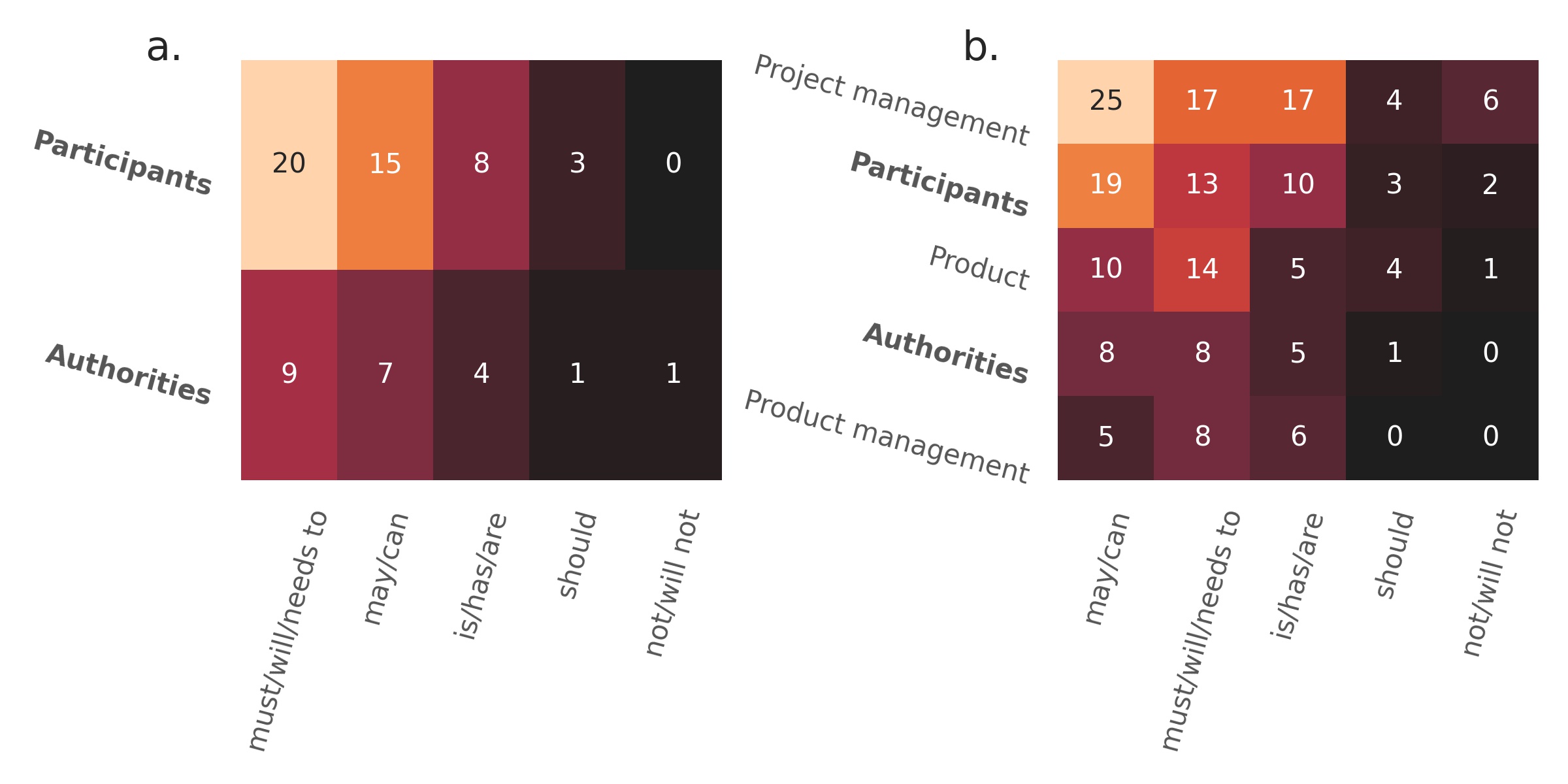}
\caption{\label{fig:crosstabs_full} Full crosstabs for visualizing interactions between Agents and Objects }
\end{figure}

\begin{figure}[hbt!]
\centering
\includegraphics[width=0.9\textwidth]{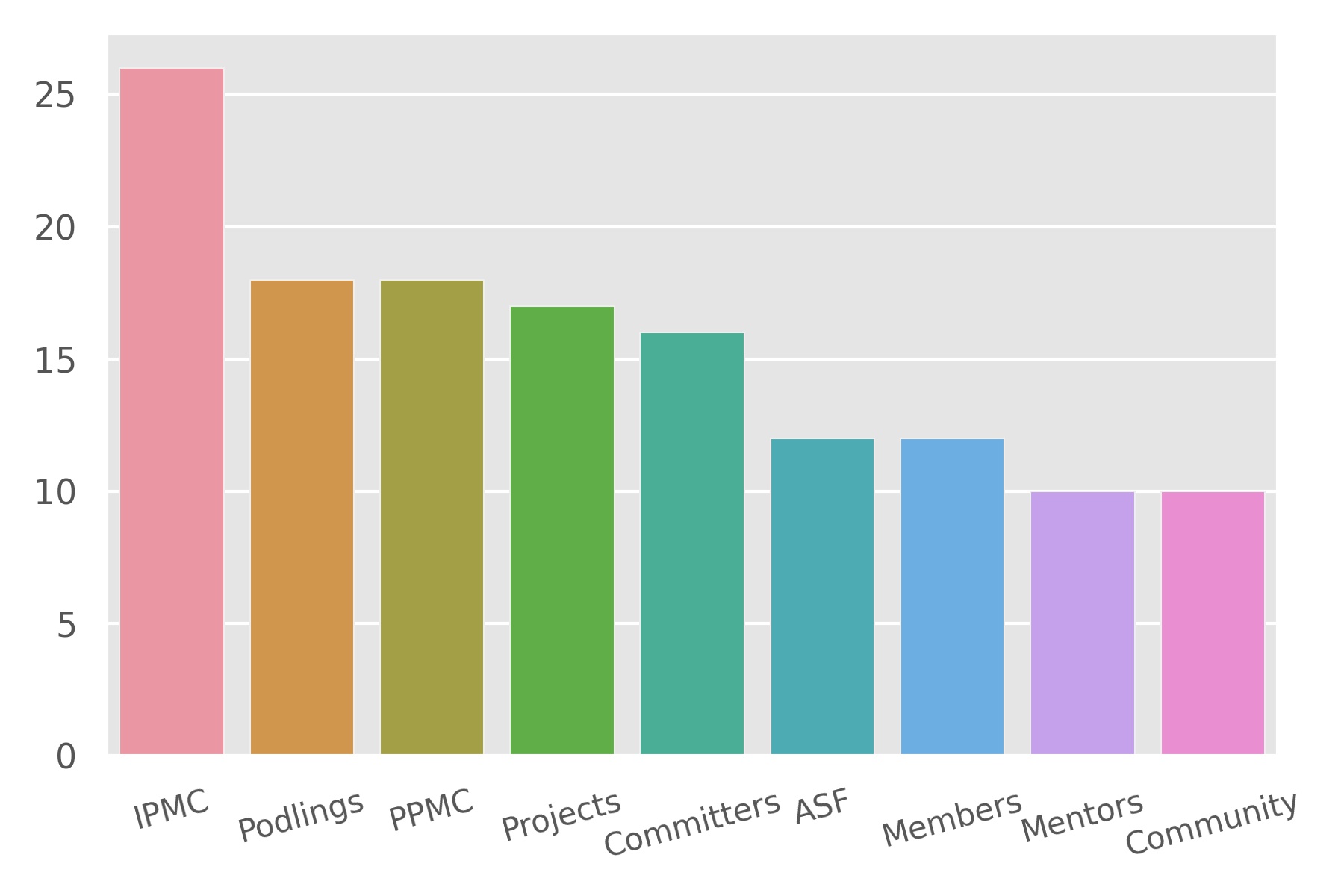}
% \caption{\label{fig:attributes_full} Key agents in the ASF }
\end{figure}

\begin{figure}[hbt!]
\centering
\includegraphics[width=0.9\textwidth]{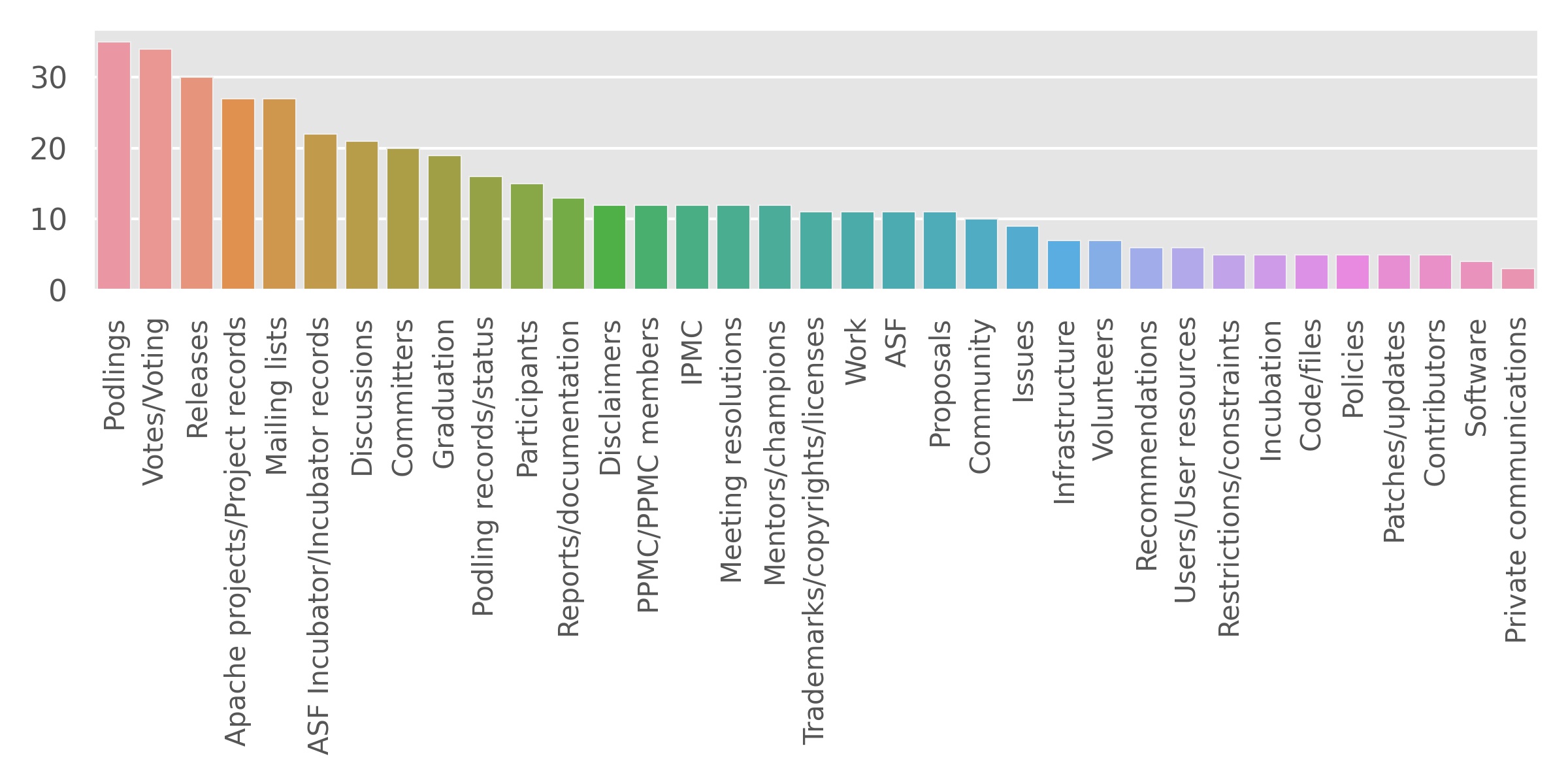}
\caption{\label{fig:objects_full} \textbf{ASF’s youngest projects (“podlings”) and their governing committees are the most regulated agents in the ASF policy documents.} Podlings are also the most likely to be the object against which another agent’s behavior is regulated, following by project management constructs such as votes, releases, and records. These two tables give the low-level agents and objects that Figures 3 and 4 were aggregated from. The first concisely represents the range of stakeholders in the ASF’s institutional ecosystem, both agents in authority roles (IPMCs, mentors, and the ASF itself) and in participant roles (the remainder). It is apparent that number of objects exceeds that of agents because objects can be animate or inanimate, but it also indicates the variety of institutional arenas that merit behavioral regulation. Specifics of the mapping are reported in \textbf{Appendix}. }
\end{figure}

\end{document}